\documentclass[aps,prd,11pt,twoside,tightenlines,nofootinbib,showpacs,preprint,superscriptaddress]{revtex4-1}
\usepackage[colorlinks=true]{hyperref}
\usepackage{xcolor}
\hypersetup{colorlinks=true, citecolor=blue, urlcolor=blue, linkcolor=blue}
\usepackage{amsmath,amssymb}
\usepackage[final]{graphicx}
\usepackage{tabularx}
\usepackage[toc,page]{appendix}
\usepackage{rotating}

\newcommand{\tabincell}[2]{\begin{tabular}{@{}#1@{}}#2\end{tabular}}

\begin{document}

\title{Study of pion vector form factor and its contribution to the muon $(g-2)$}

\author{Jing-Yu Yi}
\affiliation{School of Physics and Electronics, Hunan Key Laboratory of Nanophotonics and Devices, Central South University, Changsha 410083, China}

\author{Zhong-Yu Wang}
\affiliation{School of Physics and Electronics, Hunan Key Laboratory of Nanophotonics and Devices, Central South University, Changsha 410083, China}

\author{C. W. Xiao}
\email{xiaochw@csu.edu.cn}
\affiliation{School of Physics and Electronics, Hunan Key Laboratory of Nanophotonics and Devices, Central South University, Changsha 410083, China}

\date{\today}

\begin{abstract}

In the present work, we investigate several theoretical models of the pion vector form factor and aim at getting the best fit for the two-pion cross sections to reduce the uncertainties of the calculation of two-pion contribution to the muon anomalous magnetic moment. Combined with a polynomial description to the pion vector form factor, we obtain the best fit from the Gounaris-Sakurai (or K\"uhn-Santamaria) model for the experimental data up to 1 GeV. By product, the branching ratio of $\omega \to \pi\pi$ can be extracted as $\text{Br}(\omega\to\pi\pi)=(1.52\pm 0.06) \%$, which is consistent with the one of Particle Data Group. With the best fit of the data, we obtain the muon anomalous magnetic moment from two-pion contribution as $a_{\mu}^{\mathrm{HVP,\; LO}}(\pi^{+} \pi^{-} \leq 1\ \text{GeV}) = (497.76\pm3.15) \times 10^{-10}$. Our results are consistent with the other works.

\end{abstract}
\pacs{}

\maketitle

\section{Introduction}

The muon magnetic moment is an important and long historical issue in particle physics~\cite{Garwin:1957hc,Garwin:1960zz,Terazawa:1968jh,Terazawa:1968mx,Terazawa:1969ih,Kinoshita:1970js,Terazawa:2018pdc}. Using Dirac theory, the gyromagnetic ratio $g_\mu$ is predicted as $g_\mu=2$ for the structureless and spin $\frac{1}{2}$ muon. In fact, due to the developments of the experiments and theories, it is found that  $g_\mu$ is slightly greater than 2, which can be referred to as the anomalous magnetic moment $a_{\mu}=(g_\mu-2) / 2$. As already known, it is Schwinger's value of $a_{\mu}=\alpha / (2\pi) \simeq 0.00116$ from one loop QED radiative corrections, which is universal for all leptons. More discussions for that can be found in the reviews of~\cite{Jegerlehner:2009ry,Miller:2012opa}. 
The anomalous magnetic moment of the muon $a_{\mu}$ is experimentally and theoretically known to very high accuracy. Its measurement at Brookhaven National Laboratory was reported as~\cite{Bennett:2006fi,Mohr:2012tt} 
\begin{equation}
a_{\mu}^{\mathrm{exp}}=(11659209.1 \pm 5.4 \pm 3.3 )\times 10^{-10},
\end{equation}
where the errors were given by statistical and systematic uncertainties, respectively. The Standard Model (SM) prediction was given by~\cite{Davier:2019can}
\begin{equation}
a_{\mu}^{\mathrm{SM}}=(11659183.1 \pm 4.0 \pm 2.6 \pm 0.1)\times 10^{-10}.
\end{equation}
And thus, the difference between the experiment and theory is
\begin{equation}
\Delta a_{\mu}=a_{\mu}^{\exp }-a_{\mu}^{\mathrm{SM}}=(26.0 \pm 7.9 )\times 10^{-10},
\end{equation}
where one can see that there is a discrepancy of about 3.3$\sigma$ between the measured value and the full Standard Model prediction. But, this discrepancy has been updated with the recent results both in theory calculations and experimental measurements. The latest measurement of the anomalous magnetic moment of the muon was performed at Fermilab National Accelerator Laboratory Muon $g-2$ Experiment~\cite{Muong-2:2021hlp}, gotten
\begin{equation}
a_{\mu}^{\mathrm{exp}}=(11659204.0 \pm 5.4) \times 10^{-10}. 
\end{equation}
Combined with the measurement at Brookhaven above, one can easily get the experimental average of 
\begin{equation}
a_{\mu}^{\mathrm{exp}}=(11659206.1 \pm 4.1) \times 10^{-10}.
\end{equation}
Note that the reported results of Ref.~\cite{Bennett:2006fi} were used for this average.
The latest SM prediction was given by the recent review of the White Paper (WP)~\cite{Aoyama:2020ynm}
\begin{equation}
a_{\mu}^{\mathrm{SM}}=(11659181.0 \pm 4.3) \times 10^{-10}. 
\end{equation}
Therefore, the difference between the experiment and the theory is updated as
\begin{equation}
\Delta a_{\mu}^{\mathrm{new}}=a_{\mu}^{\exp }-a_{\mu}^{\mathrm{SM}}=(25.1 \pm 5.9) \times 10^{-10},
\end{equation}
which leads to a discrepancy of $4.2 \sigma$. This discrepancy possibly hints the new physics beyond SM and draws much theoretical attention~\cite{Chiang:2021pma,CarcamoHernandez:2021iat,Arcadi:2021cwg,Zhu:2021vlz,Endo:2021zal,Han:2021gfu,Das:2021zea,Ge:2021cjz,Baum:2021qzx,Zhang:2021gun,Ahmed:2021htr,Cao:2021tuh,Crivellin:2021rbq,Athron:2021iuf,Yin:2020afe,Yin:2021yqy,Yin:2021mls}. More discussions can be referred to Refs.~\cite{Athron:2021iuf,Terazawa} for new physics beyond SM and the latest review of~\cite{Keshavarzi:2021eqa} for recent status, and references therein.

So it is crucial to know the prediction of the SM as precisely as possible. The prediction of the SM $a_{\mu}^{\mathrm{SM}}$ can be divided into several different contributions~\cite{Aoyama:2020ynm,Zyla:2020zbs},
\begin{equation}
a_{\mu}^{\mathrm{SM}}=a_{\mu}^{\mathrm{QED}}+a_{\mu}^{\mathrm{had}}+a_{\mu}^{\mathrm{weak}},
\end{equation}
where $a_{\mu}^{\mathrm{QED}}$ is the pure electromagnetic contribution, $a_{\mu}^{\mathrm{had}}$ is the hadronic contribution, and $a_{\mu}^{\text {weak}}$ accounts for the electroweak corrections due to the exchange of the weak interacting bosons. At present,  $a_{\mu}^{\mathrm{QED}}$ was calculated with high accuracy up to five-loop order~\cite{Aoyama:2012wj,Aoyama:2012wk,Baikov:2013ula,Volkov:2019phy}, and $a_{\mu}^{\text{weak }}$ was also done up to two-loop order~\cite{Czarnecki:2002nt,Gnendiger:2013pva,Ishikawa:2018rlv}, which were given by
\begin{align}
a_{\mu}^{\mathrm{QED}} &= (11658471.8931 \pm 0.0104) \times 10^{-10}, \\
a_{\mu}^{\text{weak }} &= (15.36 \pm 0.10) \times 10^{-10},
\end{align}
where one can see the reviews of~\cite{Aoyama:2020ynm,Keshavarzi:2021eqa,Zyla:2020zbs} for more details. Thus, the large uncertainties of $a_{\mu}^{\mathrm{SM}}$ mainly come from the hadronic part of $a_{\mu}^{\mathrm{had}}$ due to the confinement and non-perturbative properties in the low energy region, which can be divided into two parts, one part from hadronic light-by-light (HLbL) scattering ($a_{\mu}^{\mathrm{HLbL}}$) and the other one from the hadronic vacuum polarization (HVP) contribution ($a_{\mu}^{\mathrm{HVP}}$). 

For the HLbL contribution~\cite{Melnikov:2003xd,Masjuan:2017tvw,Colangelo:2017fiz,Hoferichter:2018kwz,Gerardin:2019vio,Bijnens:2019ghy}, the phenomenological estimation was given by the WP~\cite{Aoyama:2020ynm}
\begin{equation}
a_{\mu}^{\mathrm{HLbL}} = (9.2 \pm 1.9) \times 10^{-10},
\end{equation}
which was consistent with Lattice QCD calculations $a_{\mu}^{\mathrm{HLbL}} =(7.87 \pm 3.06 \pm 1.77) \times 10^{-10}$~\cite{Blum:2019ugy} and $a_{\mu}^{\mathrm{HLbL}} =(10.68 \pm 1.47) \times 10^{-10}$~\cite{Chao:2021tvp} within the uncertainties. Recently (after the WP), with a model-independent method, the effects of short-distance constraints on the $a_{\mu}^{\mathrm{HLbL}}$ were evaluated in Ref.~\cite{Ludtke:2020moa} by considering the known states below 1 GeV, which obtained $a_{\mu}^{\mathrm{HLbL}} = (0.91 \pm 0.50) \times 10^{-10}$ for the contribution of pseudoscalar ground-states and $a_{\mu}^{\mathrm{HLbL}} = (0.26 \pm 0.15) \times 10^{-10}$ for the one of isovector parts. The short-distance expansion for the four-point function was derived in Ref.~\cite{Bijnens:2020xnl} via a systematic operator product expansion, where it was found that the contribution of the massless quark loop in leading order to the $a_{\mu} ^{\mathrm{HLbL}}$ was dominant and the ones from higher order were estimated to be small. In a further work of~\cite{Bijnens:2021jqo}, the perturbative QCD correction to the massless quark loop was computed, and they found that the correction up to two loops was a quite small contribution to the $a_{\mu}^{\mathrm{HLbL}}$. Ref.~\cite{Masjuan:2020jsf} discussed the short-distance constraints to the calculation of $a_{\mu}^{\mathrm{HLbL}}$ from the contribution of the axial-vector mesons. Employing resonance chiral theory, the axial-vector contribution to the $a_{\mu}^{\mathrm{HLbL}}$ was discussed in Ref.~\cite{Roig:2019reh} with a small result of $a_{\mu}^{\mathrm{HLbL;\; A}} = (0.8^{+3.5}_{-0.8})\times 10^{-11}$. On the other hand, with a warped five-dimensional model, Ref.~\cite{Cappiello:2019hwh} considered the contributions of pseudoscalar and axial-vector resonances to the $a_{\mu}^{\mathrm{HLbL}}$ and obtained a value of $a_{\mu}^{\mathrm{HLbL;\; P+A}} = (12.5\pm 1.5) \times 10^{-10}$ with much larger role for the axial-vector contribution.  In Ref.~\cite{Zanke:2021wiq}, the transition form factors of the resonance $f_1(1285)$ were analyzed in detail with the framework of vector meson dominance due to its contribution to the HLbL scattering, see more details in Ref.~\cite{Leutgeb:2019gbz}. Using dispersion relations, Ref.~\cite{Danilkin:2021icn} considered the contribution of scalar resonances to the $a_{\mu}^{\mathrm{HLbL}}$ and obtained an estimate of $a_{\mu}^{\mathrm{HLbL}} [\mathrm{scalar}] = (-0.9 \pm 0.1) \times 10^{-10}$. Moreover, several models for the short-distance constraints to the calculation of $a_{\mu}^{\mathrm{HLbL}}$ were investigated detailedly in Ref.~\cite{Colangelo:2021nkr}, where the perturbative QCD correction was also taken into account and the result of the perturbative corrections to the operator product expansion was updated as $a_{\mu}^{\mathrm{HLbL}} = (1.3 \pm 0.5) \times 10^{-10}$.

On the other hand, at the current status, the total HVP contribution was estimated as~\cite{Aoyama:2020ynm,Keshavarzi:2021eqa}
\begin{equation}
a_{\mu}^{\mathrm{HVP}} =  (684.5 \pm 4.0) \times 10^{-10},
\end{equation}
which included the leading order (LO)~\cite{Davier:2019can,Keshavarzi:2019abf}, next-to-leading order~\cite{Keshavarzi:2019abf} and next-next-to-leading order~\cite{Kurz:2014wya} contributions. In fact, the dominant one is the LO part, given by the data-driven calculations~\cite{Aoyama:2020ynm,Davier:2017zfy,Keshavarzi:2018mgv,Colangelo:2018mtw,Hoferichter:2019mqg,Davier:2019can,Keshavarzi:2019abf}
\begin{equation}
a_{\mu \quad \mathrm{DaDr}}^{\mathrm{HVP,\; LO}} =  (693.1 \pm 4.0) \times 10^{-10},
\label{eq:amuHVPLOd}
\end{equation}
which is overlapped with the lattice world average for the total LO HVP contribution~\cite{Aoyama:2020ynm},
\begin{equation}
a_{\mu \quad \mathrm{lat}}^{\mathrm{HVP,\; LO}} =  (711.6 \pm 18.4) \times 10^{-10}.
\end{equation}
But, the recent calculation of lattice QCD for the LO HVP contribution was reported as~\cite{Borsanyi:2020mff}
\begin{equation}
a_{\mu \quad \mathrm{lat}}^{\mathrm{HVP,\; LO}}= (707.5 \pm 5.5) \times 10^{-10}, 
\end{equation}
with high accuracy, which is a bit smaller than the one obtained in Ref.~\cite{Lehner:2020crt}, $(714 \pm 27 \pm 13) \times 10^{-10}$. These new lattice results lead the SM prediction to be in agreement with the current experimental measurement and the new physics to be questionable.

One thing should be mentioned that, the LO part from the data-driven calculations in Eq.~\eqref{eq:amuHVPLOd} is only taken the $e^+ e^-$ annihilation data into account, since the one from $\tau$ decay data is not precise enough at present. With the results of Ref.~\cite{Davier:2013sfa}, there was still $2.2\; \sigma$ discrepancy between the $e^+ e^-$ based and $\tau$  based results. Recently, with resonance chiral theory supplemented by dispersion relations, Ref.~\cite{Gonzalez-Solis:2019iod} studied the pion vector form factor using the experimental data of $\tau$ decay from Belle and recent BaBar measurements. Based on these results, the $a_{\mu}^{\mathrm{HVP,\; LO}}$ was extracted from the $\tau$ decay data of $\tau^- \to \pi^- \pi^0 \nu_\tau$ in a further work of~\cite{Miranda:2020wdg}, obtained $a_{\mu \quad \tau\;\mathrm{Dat}}^{\mathrm{HVP,\; LO}} =  (705.7^{+4.0}_{-4.1}) \times 10^{-10}$ and $a_{\mu \quad \tau\;\mathrm{Dat}}^{\mathrm{HVP,\; LO}} =  (700.7^{+6.1}_{-5.2}) \times 10^{-10}$ for different strategies. Using a framework of the hidden local symmetry model combined with appropriate symmetry breaking mechanisms, both the $e^+ e^-$ annihilation and $\tau$ decay data were analysed in Ref.~\cite{Benayoun:2019zwh}, where a value of $a_{\mu}^{\mathrm{HVP,\; LO}} =  (686.65 \pm 3.01) \times 10^{-10}$ with the uncertainties of $\rho-\gamma$ mixing was reported and a further result was updated in the recent work of~\cite{Benayoun:2021ody}. 

As one can see that, for the SM prediction $a_{\mu}^{\mathrm{SM}}$, the hadronic part $a_{\mu}^{\mathrm{had}}$ still has large uncertainties, especially for the one $a_{\mu}^{\mathrm{HVP}}$. Recently, to solve the inverse problem to the dispersion relation, a value for the HVP contribution was obtained as $a_{\mu}^{\mathrm{HVP}} = (641^{+65}_{-63}) \times 10^{-10}$ in Ref.~\cite{Li:2020fiz}. Based on the chiral perturbation theory, Ref.~\cite{Aubin:2020scy} discussed that the finite-volume corrections to $a_{\mu}^{\mathrm{HVP}}$ could be precisely evaluated, where once all low-energy constants were already known. Ref.~\cite{Malaescu:2020zuc} investigated the potential impact on the electroweak fits of the tensions between the current determinations of the HVP contributions to the $a_{\mu}$, based on either phenomenological calculations or Lattice QCD calculations. Note that, taking into account the measurement of the Higgs mass, the impact of HVP on $a_{\mu}^{\mathrm{SM}}$ and the global fits to electroweak precision data was studied in Ref.~\cite{Crivellin:2020zul}, where some options for physics beyond SM were discussed. 

To reduce the uncertainties of the part $a_{\mu}^{\mathrm{HVP}}$, indeed, the accurate evaluations must rely on the corresponding cross section measurements for the normal calculations of data-driven. Combined with the Effective Lagrangian, an iterated global fit scheme was adopted in Ref.~\cite{Benayoun:2015gxa} to reduce the uncertainties for the description of the $e^+ e^- \to \pi^+ \pi^-$ annihilation data up to 1.05 GeV. With a dispersive representation of the pion vector form factor (PVFF), different constraints on the two-pion contribution were examined for its effects on the $a_{\mu}^{\mathrm{HVP}}$ in Ref.~\cite{Colangelo:2020lcg}, where a value of $a_{\mu}^{\mathrm{HVP}} (\pi\pi \leq 1 \text{GeV}) = 497.0(1.4) \times 10^{-10}$ was gotten in one case of their fits. Using a parametrization-free formalism based on analyticity and unitarity, the PVFF and its contribution to the $a_{\mu}^{\mathrm{HVP}}$ were investigated in Refs.~\cite{Ananthanarayan:2016mns,Ananthanarayan:2018nyx,Ananthanarayan:2020vum} for the energy range around the $\rho(770)$ resonance. As already known, about $73\%$ of the LO hadronic contribution and about $60\%$ of the total uncertainty are given by $e^+ e^-$ annihilated to the $\pi^{+} \pi^{-}(\gamma)$ final states, which are dominated by the $\rho(770)$ resonance. Therefore, it is important to study the $e^+ e^- \to \pi^+ \pi^-$ annihilation, which always relates to the PVFF. Thus, in the present work, we study several theoretical models of the PVFF, and aim at finding out the best fit of the $\pi^{+} \pi^{-}$ scattering cross sections. In the next section, we first introduce the calculation of $a_{\mu}^{\mathrm{HVP,\; LO}}$ with data-driven briefly. Following, we discuss several phenomenological models of the PVFF, combined with a polynomial description, or equivalently how to take into account the contribution of the $\rho(770)$ resonance. Then, we obtain the results from fitting the PVFF data of the collaborations Orsay, DM1, OLYA, CMD1, CMD2, BABAR, BESIII, KLOE, SND, and so on. And thus, we perform a calculation of $a_{\mu}^{\mathrm{HVP,\; LO}}$ up to 1 GeV. At the end, it is our conclusion.

\section{Muon ($g-2$) calculation with dispersion relation}

As we discussed above, the theoretical prediction of $a_{\mu}^{\mathrm{SM}}$ has large uncertainties from the parts of hadronic contribution $a_{\mu}^{\mathrm{had}}$. Due to the confinement, the non-perturbative properties become dominant in the low energy region, where the quarks are confined inside hadrons. Therefore, perturbative QCD fails to evaluate the hadronic (quark and gluon) loop contributions to the $a_{\mu}^{\mathrm{SM}}$ precisely. In principle, one can do the calculation of $a_{\mu}^{\mathrm{had}}$ from first principle calculation in lattice QCD. But, most of the evaluations in lattice QCD are still not precise enough, except for the recent one of~\cite{Borsanyi:2020mff}. Alternatively, the HVP contributions $a_{\mu}^{\mathrm{HVP}}$ can be calculated with the data-driven approach, which uses the dispersion relation together with the optical theorem and experimental data. Thus, the LO HVP contribution to the $a_{\mu}^{\mathrm{SM}}$ can be calculated via a dispersion relation using the measured cross sections of $e^{+} e^{-} \rightarrow$ hadrons~\cite{Gourdin:1969dm}
\begin{equation}
a_{\mu}^{\mathrm{HVP,\; LO}}=\frac{1}{\pi} \int_{0}^{\infty} \frac{\mathrm{d} s}{s} \operatorname{Im} \Pi^{(\mathrm{H})}(s) K(s), 
\end{equation} 
where the kernel function is given by
\begin{equation}
K(s)=\left(\frac{\alpha}{\pi}\right)\left\{\frac{1}{2} x^{2}\left(2-x^{2}\right)+(1+x)^{2}\left(1+x^{2}\right) \frac{\ln (1+x)-x+\frac{1}{2} x^{2}}{x^{2}}+\frac{1+x}{1-x} x^{2} \ln x\right\}, 
\end{equation}
with the definitions
\begin{equation}
\beta_{\mu}=\sqrt{1-\left(4 m_{\mu}^{2} / s\right)} \text { , } x=\left(1-\beta_{\mu}\right) /\left(1+\beta_{\mu}\right),
\end{equation}
and the electromagnetic coupling taking as $\alpha=e^{2} /(4 \pi) \approx 1/137.036$ from Particle Data Group (PDG)~\cite{Zyla:2020zbs}. Note that $s$ is the total energy of two-body system, $s\equiv (p_1 + p_2)^2$. With the optical theorem, the imaginary part of the vacuum polarization amplitude $\operatorname{Im}\mathrm{\Pi }^{(\mathrm{H})}(s)$ can be expressed in terms of the total cross section of the electron-positron annihilation into hadrons
\begin{equation}
\sigma_{\text {tot}}\left(\mathrm{e}^{+} \mathrm{e}^{-} \rightarrow \text{hadrons }\right)=\frac{4 \pi^{2} \alpha}{s} \frac{1}{\pi} \operatorname{Im} \Pi^{(\mathrm{H})}(s).
\end{equation}
Thus, one can deduce
\begin{equation}
a_{\mu }^{\mathrm{HVP,\; LO}}=\frac{1}{4 \pi^{2} \alpha} \int_{m_{\pi}^{2}}^{\infty} \mathrm{d} s \; \sigma_{\text{tot}}\left(\mathrm{e}^{+} \mathrm{e}^{-} \rightarrow \text{hadrons }\right) K(s),
\label{eq:amuHVPLO}
\end{equation}
which uses the measured bare cross sections for the annihilation $\mathrm{e}^{+} \mathrm{e}^{-} \rightarrow $ hadrons as inputs, and where the lower limit of the dispersion integral is in fact the $\pi^0\,\gamma$ cut. One should keep in mind that the experimentally measured cross sections are the dressed one, where the bare cross sections can be corrected by the running of the coupling constant $\alpha(s)$. In fact, this correction has always been done in the experimental data reported. Thus, the corresponding cross section measurements play a key role in the accurate evaluation of $a_{\mu }^{\mathrm{had,\; LO}}$.

Note that, the kernel function $K(s)$ decreases monotonically with increasing $s$, so it gives strong weight to the low-energy part of the integral, where about $73\%$ of the LO hadronic contributions are given by the $\pi^{+} \pi^{-}(\gamma)$ final states, dominated by the $\rho(770)$ resonance. In the present work, we focus on the energy region of about 1 GeV, which is mainly contributed by the $\pi^{+} \pi^{-}(\gamma)$ final states. The total cross section contributed by two-pion final states is given by
\footnote{In fact, the two-pion cross section is inclusive of final state radiation effects and exclusive of all vacuum polarization effects in the experimental measurements.}
\begin{equation}
\sigma_{\mathrm{tot}}\left(\mathrm{e}^{+} \mathrm{e}^{-} \rightarrow \pi^{+} \pi^{-}\right)=\frac{1}{3} \pi \alpha^{2} \frac{1}{s} \beta_{\pi}^3(s) \mid F_{\pi}(s)\mid^{2},
\label{eq:croFF}
\end{equation}
where $F_{\pi}(s)$ is the PVFF and the pion phase is defined as $\beta_{\pi}(s)=\sqrt{1-4 m_{\pi}^{2} / s}$. Then, it is important to study the model of PVFF, see the discussion in the next section. Thus, the two-pion contribution to the anomalous magnetic moment of the muon can be written as
\begin{equation}
a_{\mu}^{\mathrm{HVP,\; LO}}(\pi^{+} \pi^{-})=\frac{\alpha}{12 \pi} \int_{4 m_{\pi}^{2}}^{\infty} \frac{\mathrm{d} s}{s} \beta_{\pi}^3(s) \left|F_{\pi}(s)\right|^{2} K(s).
\label{eq:amupi}
\end{equation}

\section{The model for the pion form factor}

In the present work, we are interested in the experimental data at a centre-of-mass energy below 1 GeV, which is around the energy region corresponding to the $\rho$ resonance. For the cross section of $\mathrm{e}^{+} \mathrm{e}^{-} \rightarrow \pi^{+} \pi^{-}$, it can be associated with the PVFF, see Eq.~\eqref{eq:croFF}, which can be defined as
\begin{equation}
\langle \pi^+(p^\prime) \pi^-(p)| J_\mu(0) | 0 \rangle = (p^\prime -p)_\mu F_\pi (s),
\end{equation}
with $s=(p^\prime +p)^2$ and $J_\mu$ the vector-isovector current. For our case of energy range below 1 GeV, in terms of the pion $P$-wave phase shift $\delta_{11}(s)$, the PVFF fulfills the following discontinuity condition,
\begin{equation}
\text{Im}\; F_\pi (s) = F_\pi (s) \sin \delta_{11}(s) e^{-i \delta_{11}(s)} \Theta(s-4 m_\pi^2).
\label{eq:disfv}
\end{equation}
With a once subtracted dispersion relation, the solution of Eq.~\eqref{eq:disfv} can be written into a general ansatz~\cite{Roos:1975zf,Lang:1975ge}
\begin{equation}
F_{\pi}(s)=P(s) \Omega(s), 
\label{eq:fv}
\end{equation}
where $P(s)$ is a polynomial and $\Omega(s)$ is the Omn\'es function~\cite{Omnes:1958hv}. Note that the solution for higher subtracted dispersion relation can be referred to Refs.~\cite{Pich:2001pj,Hoferichter:2014vra,Isken:2017dkw} for more discussions and applications, where especially in Ref.~\cite{Pich:2001pj} the experimental data of the PVFF up to $\sqrt{s} \simeq 1.2$ GeV was in a good description with two subtraction constants of thrice-subtracted dispersion relation and the two-pion contribution to the $a_{\mu}^{\mathrm{HVP,\; LO}}$ was evaluated. Furthermore, this ansatz was extrapolated to the radiative decays of $\eta^{(\prime)} \to \pi^+ \pi^- \gamma$~\cite{Stollenwerk:2011zz,Hanhart:2013vba}, where a linear polynomial was used,
\begin{equation}
P(s)=1+ \alpha s,
\end{equation}
with $\alpha$ a free parameter, determined from the data. Indeed, the linear behaviour was clearly shown in the results of Refs.~\cite{Stollenwerk:2011zz,Hanhart:2013vba} for the data of the PVFF below 1 GeV. A new parameterization to the PVFF for a full energy range can be referred to Ref.~\cite{Hanhart:2012wi}, where the isospin violation mechanism was also considered, such as the mixing effects of $\rho-\omega$ and $\omega-\phi$. Besides, the Omn\'es function is given by
\begin{equation}
\Omega(s)=\exp \left(\frac{s}{\pi} \int_{4m_{\pi}^{2}}^{\infty} \frac{d s^{\prime}}{s^{\prime}} \frac{\delta_{11}\left(s^{\prime}\right)}{s^{\prime}-s-i\epsilon}\right),
\label{eq:omn}
\end{equation}
where the phase shift ${\delta_{11}\left(s\right)}$ can be taken from the Madrid model's results~\cite{GarciaMartin:2011cn}. In Ref.~\cite{GarciaMartin:2011cn}, the phase shift for $s^{1 / 2} \leq 2 m_{K}$ fulfilled
\begin{equation}
\begin{aligned}
\cot \delta_{11}(s)&=\frac{s^{1 / 2}}{2 k^{3}}\left(m_{\rho}^{2}-s\right)\left\{\frac{2 m_{\pi}^{3}}{m_{\rho}^{2} \sqrt{s}}+B_{0}+B_{1} w(s)\right\},\\
w(s)&=\frac{\sqrt{s}-\sqrt{s_{0}-s}}{\sqrt{s}+\sqrt{s_{0}-s}}, \quad s_{0}^{1 / 2}=1.05  \mathrm{ GeV},
\end{aligned}
\end{equation}
where the $\rho$ mass was fixed to ${m_{\rho}}=773.6$ MeV, the other masses were taken as ${m_{\pi}}=139.57 $ MeV, ${m_{K}}=496$ MeV,  ${m_{\eta}}=547.51$ MeV, and the central-mass momentum in the two-pion final states $k=\sqrt{s-4 m_{\pi}^{2}}/2$. For $2 m_{K} \leq s^{1 / 2} \leq 1420 $ MeV, one can have
\begin{equation}
\delta_{11}(s)=\lambda_{0}+\lambda_{1}\left(\sqrt{s} / 2 m_{K}-1\right)+\lambda_{2}\left(\sqrt{s} / 2 m_{K}-1\right)^{2},
\end{equation}
where $\lambda_{0}$ is fixed from the value of $\delta_{11}(4 m_{K}^{2})$ obtained from the low energy parametrization, so that the phase shift is continuous. Besides, the parameters of $B_{0},B_{1},\lambda_{1}, \lambda_{2}$ were determined with UFD set or CFD set in Ref.~\cite{GarciaMartin:2011cn}. For higher energy, we choose the phase shifts close to $\pi$ smoothly, written as
\begin{equation}
\delta_{11}(s)=\pi-\frac{a}{b-s},
\end{equation}
where the coefficients $a$ and $b$ are determined with the value of $\delta_{11}(s_{0})$ ($\sqrt{s_{0}}=1420 $ MeV for example, in fact, we take $\sqrt{s_{0}}=1300 $ MeV for the best results) to make it continuous, and also keep the derivative at $s=s_{0}$.

In fact, as discussed above, the data for the cross section of two pions below 1 GeV is dominated by the contribution of the $\rho$ resonance
\footnote{Note that a new expression for the PVFF was proposed in Ref.~\cite{Achasov:2012bz}, where the contributions from the loops of $\pi^+\pi^-$ and $K\bar{K}$ and higher $\rho$ resonances were considered, and which described the data well in the range from -10 GeV to 1 GeV and was extrapolated to the energy up to 3 GeV in a further work of~\cite{Achasov:2013usa}.}. 
Indeed, the Omn\'es function $\Omega(s)$ in Eq.~\eqref{eq:fv} is mainly contributed by the pion p-wave phase shift, see Eq.~\eqref{eq:omn}. In the present work, we investigate how to include the contribution of $\rho$ to get better description of the experimental data. Thus, we want to know the effects of different models for the part of Omn\'es function. Note that the energy region of 1 GeV is safely below the inelastic threshold, see the discussions in Ref.~\cite{Danilkin:2014cra}. First, we use the model of Heyn and Lang (HL)~\cite{Heyn:1980bh},
\begin{equation}
\Omega(s)=\frac{c+m_{\pi}^{2} g(0)}{\widetilde{s}_{p}} \frac{\tilde{s}_{p}-s}{a s^{2}+b s+c-\left(s-4 m_{\pi}^{2}\right) g(s) / 4},
\end{equation}
where the function $g(s)$ is given by the one-pion-loop diagram in the self-energy of $\rho$ resonance,
\begin{equation}
\begin{aligned}
g(s) &=-\frac{1}{\pi} u \ln \frac{1+u}{1-u}+i u, \quad u=\sqrt{1-4 m_{\pi}^{2} / s}, \quad s \geqslant 4 m_{\pi}^{2} \\
&=-\frac{2}{\pi} u \arctan \frac{1}{u}, \quad u=\sqrt{4 m_{\pi}^{2} / s-1}, \quad 0 \leqslant s \leqslant 4 m_{\pi}^{2} \\
&=-\frac{1}{\pi} u \ln \frac{u+1}{u-1}, \quad u=\sqrt{1-4 m_{\pi}^{2} / s}, \quad s<0 ,\\
g(0) &=-2 / \pi,
\end{aligned}
\end{equation}
and the parameters $a,\ b,\ c$ are free. Besides, $\tilde{s}_{p}$ is the value of the zero of the denominator and can be determined from $a,\ b,\ c$, using the condition $f(s) = 0$, reading
\begin{equation}
f(s)=a s^{2}+b s+c-\left(s-4 m_{\pi}^{2}\right) g(s) / 4 = 0. 
\label{eq:fs}
\end{equation}

Second, as discussed above, the Omn\'es function is mainly considered the resonance contribution of $\rho$. Thus, using the vector meson dominance approach, one can replace the Omn\'es function with the simple Breit-Wigner (BW) form (BW1),
\begin{equation}
\Omega(s) \rightarrow BW1=-\frac{M_{\rho}^{2}}{s-M_{\rho}^{2}+i M_{\rho} \Gamma_{\rho}},
\end{equation}
with $M_{\rho}$ and $\Gamma_{\rho}$ as free parameters for the $\rho$ meson, which are also fitted by the experimental data. Furthermore, due to the large $\rho$ width, one can use the more common one (BW2)~\cite{Danilkin:2014cra},
\begin{equation}
\Omega(s) \rightarrow BW2=-\frac{M_{\rho}^{2}}{s-M_{\rho}^{2}-i \sqrt{s} \Gamma(s)},
\end{equation}
where the energy dependent decay width is given by
\begin{equation}
\Gamma(s)=\Gamma_{\rho}\left[\frac{p(s)}{p\left(M_{\rho}^{2}\right)}\right]^{3} \frac{M_{\rho}^{2}}{s},
\end{equation}
with the pion momentum
\begin{equation}
p(s)=\frac{1}{2} \sqrt{s} \beta_{\pi}(s)= \frac{\sqrt{s-4 m_{\pi}^{2}}}{2}.
\end{equation}

Third, for taking into account more precisely the finite-width corrections, one can also use the method of Gounaris and Sakurai (GS)~\cite{Gounaris:1968mw}, written as
\begin{equation}
\Omega(s) \rightarrow BW^{\mathrm{GS}}(s)=\frac{-\left(M_{\rho}^{2}+d M_{\rho} \Gamma_{\rho}\right)}{s-M_{\rho}^{2}-\Gamma_{\rho}\left(M_{\rho}^{2} / p_{\rho}^{3}\right)\left[p^{2}\left(h-h_{\rho}\right)-\left(s-M_{\rho}^{2}\right) p_{\rho}^{2} h_{\rho}^{\prime}\right]+i M_{\rho} \Gamma_{\rho}(s)},
\end{equation}
where
\begin{equation}
\begin{array}{l}
p=\left(s / 4-m_{\pi}^{2}\right)^{1 / 2}, \quad h(s)=\frac{2 p}{\pi \sqrt{s}} \ln \left(\frac{\sqrt{s}+2 p}{2 m_{\pi}}\right), \quad s \geqslant 4 m_{\pi}^{2}, \\
p=i\left(m_{\pi}^{2}-s / 4\right)^{1 / 2}, \quad h(s)=\frac{2 p i}{\pi \sqrt{s}} \operatorname{arccot}\left(\frac{s}{4 m_{\pi}^{2}-s}\right)^{1 / 2}, \quad 0 \leqslant s<4 m_{\pi}^{2},
\end{array}
\end{equation}
\begin{equation}
p_{\rho}=p\left(M_{\rho}^{2}\right), \quad h_{\rho}=h\left(M_{\rho}^{2}\right), \quad \Gamma_{\rho}(s)=\Gamma_{\rho}\left(\frac{p}{p_{\rho}}\right)^{3} \frac{M_{\rho}}{\sqrt{s}}, h_{\rho}^{\prime}=\left.h^{\prime}(s)\right|_{s=M_{\rho}^{2}} ,
\end{equation}
and $d$ is fixed in terms of the masses $M_{\rho}$ and $m_{\pi}$, 
\begin{equation}
d=\frac{3}{\pi} \frac{m_{\pi}^{2}}{p_{\rho}^{2}} \ln \left(\frac{M_{\rho}+2 p_{\rho}}{2 m_{\pi}}\right)+\frac{M_{\rho}}{2 \pi p_{\rho}}-\frac{m_{\pi}^{2} M_{\rho}}{\pi p_{\rho}^{3}}.
\end{equation}

Note that, in this GS model, one can also determine the $M_{\rho}$ and $\Gamma_{\rho}$ from the fits. Thus, in BaBar's paper~\cite{Lees:2012cj}, its form was changed equivalently as
\begin{equation}
BW^{\mathrm{GS}}\left(s, M_{\rho}, \Gamma_{\rho}\right)=\frac{M_{\rho}^{2}\left[1+d(M_{\rho}) \Gamma_{\rho} / M_{\rho} \right]}{M_{\rho}^{2}-s+f\left(s, M_{\rho}, \Gamma_{\rho}\right)-i M_{\rho} \Gamma\left(s, M_{\rho}, \Gamma_{\rho}\right)},
\end{equation}
where 
\begin{equation}
\Gamma\left(s, M_{\rho}, \Gamma_{\rho}\right)=\Gamma_{\rho} \frac{s}{M_{\rho}^{2}}\left[\frac{\beta_{\pi}(s)}{\beta_{\pi}\left(M_{\rho}^{2}\right)}\right]^{3}, 
\end{equation}
\begin{equation}
d\left(M_{\rho}\right)=\frac{3}{\pi} \frac{m_{\pi}^{2}}{k^{2}\left(M_{\rho}^{2}\right)} \ln \left[\frac{M_{\rho}+2 k\left(M_{\rho}^{2}\right)}{2 m_{\pi}}\right]+\frac{M_{\rho}}{2 \pi k\left(M_{\rho}^{2}\right)}-\frac{m_{\pi}^{2} M_{\rho}}{\pi k^{3}\left(M_{\rho}^{2}\right)},
\end{equation}
\begin{equation}
f\left(s, M_{\rho}, \Gamma_{\rho}\right)=\frac{\Gamma_{\rho} M_{\rho}^{2}}{k^{3}\left(M_{\rho}^{2}\right)}\left[k^{2}(s)\left(h(s)-h\left(M_{\rho}^{2}\right)\right)+\left(M_{\rho}^{2}-s\right) k^{2}\left(M_{\rho}^{2}\right) h^{\prime}\left(M_{\rho}^{2}\right)\right],
\end{equation}
with the pion phase $\beta_{\pi}(s)$ defined as in the last section, and 
\begin{equation}
k(s)=\frac{1}{2} \sqrt{s} \beta_{\pi}(s) ,
\end{equation}
\begin{equation}
h(s)=\frac{2}{\pi} \frac{k(s)}{\sqrt{s}} \ln \left(\frac{\sqrt{s}+2 k(s)}{2 m_{\pi}}\right),
\end{equation}
and $h^{\prime}(s)$ is the derivative of $h(s)$.

Moreover, similar to GS model, there is another form presented by K\"uhn and Santamaria (KS)~\cite{Kuhn:1990ad},
\begin{equation}
B W^{KS}(s)=\frac{M_{\rho}^{2}+\Gamma_{\rho} M_{\rho} d^\prime}{M_{\rho}^{2}-s+H(s)-i \sqrt{s} \Gamma_{\rho}(s)},
\end{equation}
where
\begin{equation}
H(s)=\hat{H}(s)-\hat{H}\left(M_{\rho}^{2}\right)-\left(s-M_{\rho}^{2}\right) \hat{H}^{\prime}\left(M_{\rho}^{2}\right),
\end{equation}
\begin{equation}
\hat{H}(s)=\frac{\Gamma_{\rho} M_{\rho}^{2}}{p_{\rho}^{3}}\left(s / 4-m_{\pi}^{2}\right) h(s),
\end{equation}
\begin{equation}
h\left(s\right)=\left\{\begin{array}{l}
\frac{1}{2 \pi}\left(1-\frac{4 m_{\pi}^{2}}{s}\right)^{1 / 2} \ln \left(\frac{1+\left(1-\frac{4 m_{\pi}^{2}}{s}\right)^{1 / 2}}{1-\left(1-\frac{4 m_{\pi}^{2}}{s}\right)^{1 / 2}}\right) ,4 m_{\pi}^{2} \leq s\\
\frac{i}{2 \pi}\left(\frac{4 m_{\pi}^{2}}{s}-1\right)^{1 / 2} \ln \left(\frac{i\left(\frac{4 m_{\pi}^{2}}{s}-1\right)^{1 / 2}+1}{i\left(\frac{4 m_{\pi}^{2}}{s}-1\right)^{1 / 2}-1}\right),0 \leq s \leq 4 m_{\pi}^{2}
\end{array}\right.
\end{equation}
\begin{equation}
d^\prime=\frac{3}{2 \pi} \frac{m_{\pi}^{2}}{p_{\rho}^{2}} \ln \left(\frac{M_{\rho}+2 p_{\rho}}{M_{\rho}-2 p_{\rho}}\right)+\frac{M_{\rho}}{2 \pi p_{\rho}}-\frac{m_{\pi}^{2} M_{\rho}}{\pi p_{\rho}^{3}},
\end{equation}
\begin{equation}
p(s)=\frac{1}{2}\left(s-4 m_{\pi}^{2}\right)^{1 / 2}, \quad p_{\rho}=\left(M_{\rho}^{2}-4 m_{\pi}^{2}\right)^{1 / 2} / 2.
\end{equation}

Finally, since $e^{+} e^{-}$ annihilation data has the $\rho-\omega$ mixing effects, we should take into account this effect as done in Refs.~\cite{Davier:2019can,Hanhart:2016pcd}
\begin{equation}
P(s)=1+\alpha s+\frac{\kappa s}{m_{\omega}^{2}-s-i m_{\omega} \Gamma_{\omega}},
\label{eq:ps}
\end{equation}
where we take $m_{\omega}=782.65$ MeV and $\Gamma_{\omega}=8.49$ MeV from PDG~\cite{Zyla:2020zbs}, and $\kappa$ is a free parameter containing the information of $\omega\pi\pi$ coupling, see our results later, where more discussions can be referred to Ref.~\cite{Hanhart:2016pcd}. 
Note that, Eq.~\eqref{eq:ps} fulfils $P(0)=1$, which guarantees the condition $F_{\pi}(0)=1$, except for the BW1 model. Indeed, the BW1 model is the typical one of the vector meson dominance, which is known to violate the condition~\cite{OConnell:1995nse}. Furthermore, in general, one can also float the parameters $m_{\omega}$ and $\Gamma_{\omega}$ in Eq.~\eqref{eq:ps} as done in Refs.~\cite{Hanhart:2016pcd,Davier:2019can}. But, as found in Ref.~\cite{Davier:2019can}, a value of $m_{\omega}=(782.0 \pm 0.1)$ MeV was obtained in the fitting results with all the experimental data, which is not much different from the PDG one we used. Due to the small width of the $\omega$ meson and the narrow energy region of $\rho-\omega$ mixing, we fixed them, where in fact the errors of $m_{\omega}$ and $\Gamma_{\omega}$ only contributed tiny influences to the uncertainties of final results as found later. Another thing should be mentioned that, the parameters $M_{\rho}$ and $\Gamma_{\rho}$ are model dependent. Thus, the differences between these models can be seen from our results in the next section, see more discussions later.

\section{Results}

\begin{figure}[ht] 
\centering
\includegraphics[scale=0.5]{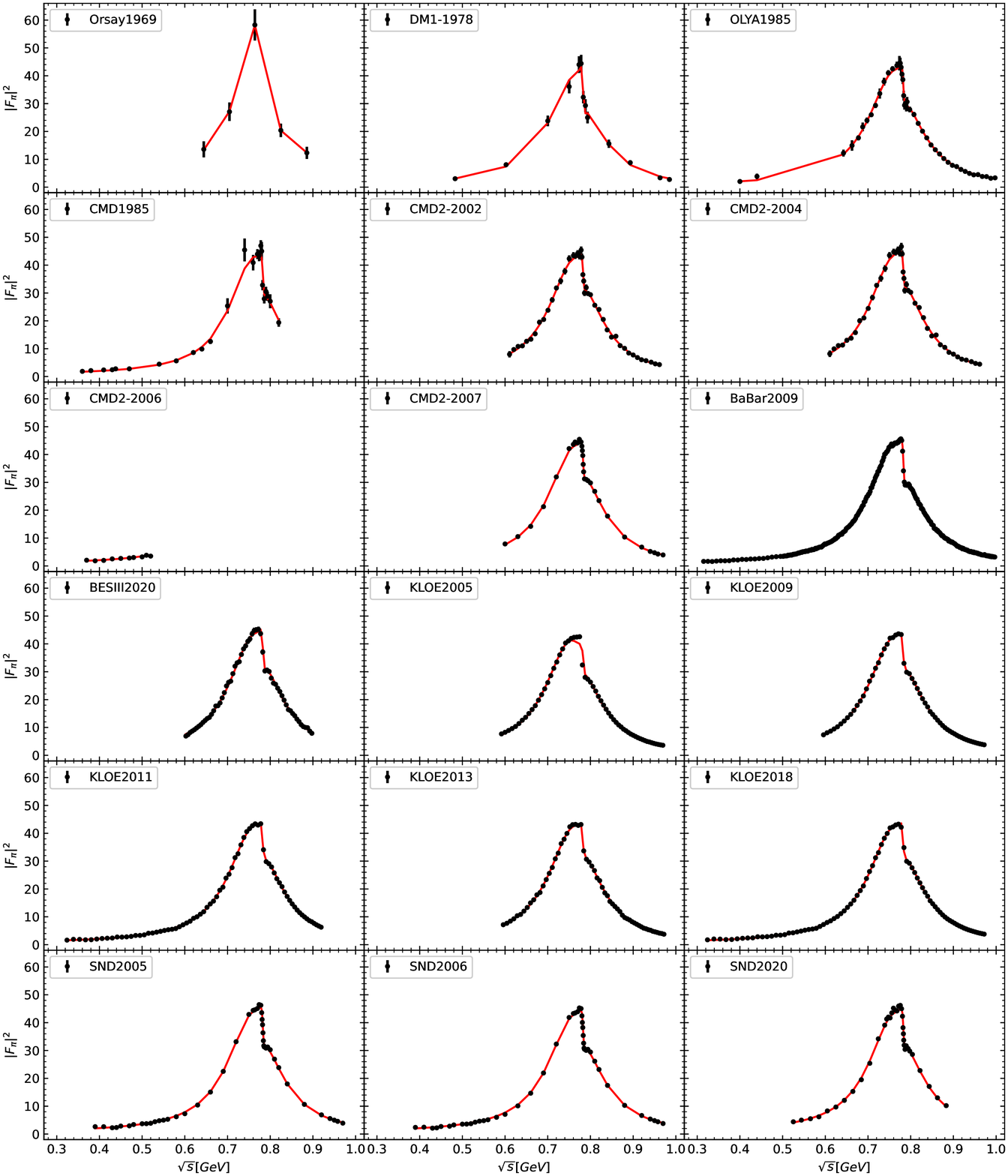}
\caption{Fitting results with the GS model for each set of experimental data.}
\label{fig:gs1}
\end{figure}

\begin{figure}[ht] 
\centering
\includegraphics[scale=1]{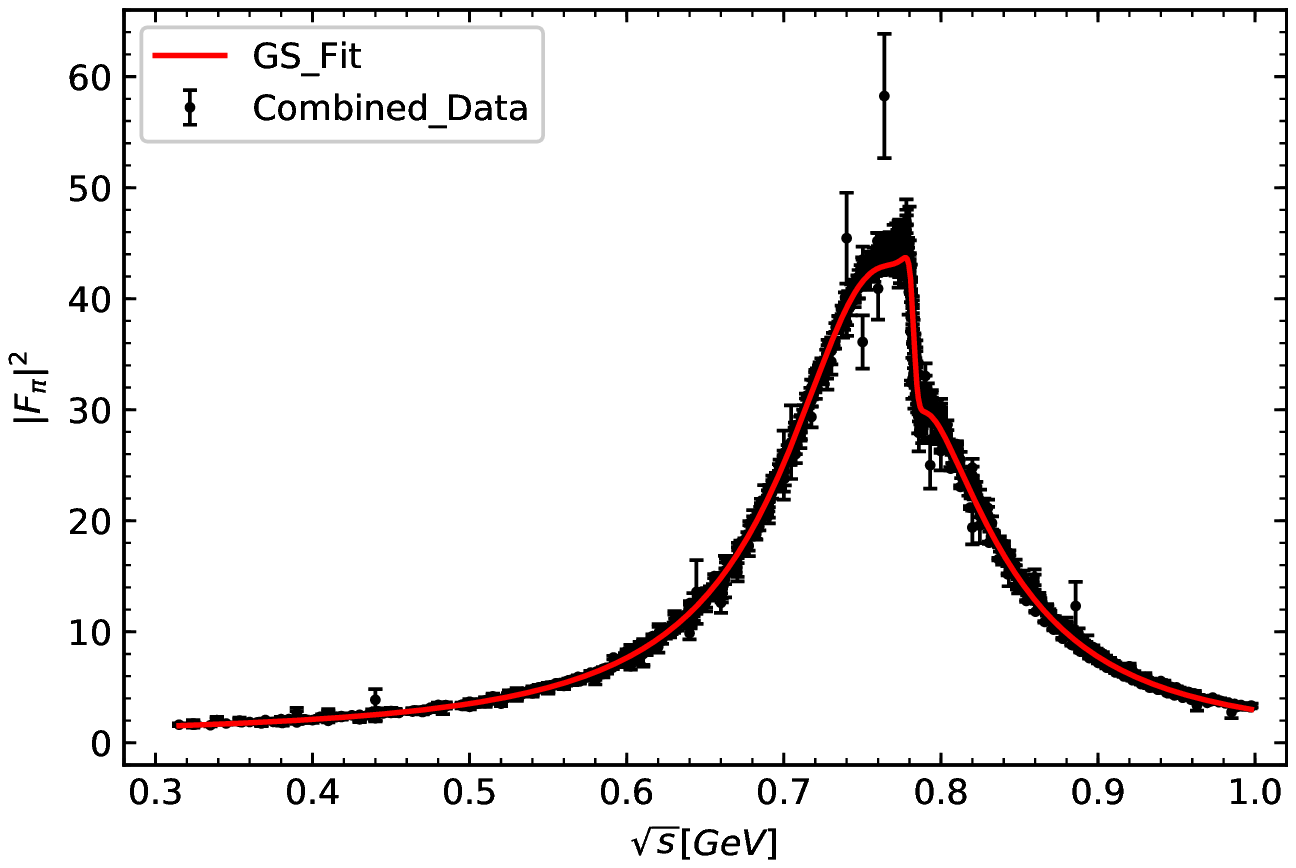}
\caption{Fitting results with the GS model for the Com. Dat..}
\label{fig:gs2}
\end{figure}

For the data of the PVFF below the energy region of 1 GeV, we take them from the experimental collaborations of Orsay, DM1, OLYA, CMD1, CMD2, BABAR, BESIII, KLOE and SND~\cite{Augustin:1969kn,Quenzer:1978qt,Barkov:1985ac,Akhmetshin:2001ig,Akhmetshin:2003zn,Aulchenko:2006na,Akhmetshin:2006wh,Akhmetshin:2006bx,Aubert:2009ad,Ablikim:2015orh,Ablikim:2020bah,Aloisio:2004bu,Ambrosino:2008aa,Ambrosino:2010bv,Babusci:2012rp,Anastasi:2017eio,Achasov:2005rg,Achasov:2006vp,Achasov:2020iys}. Our fitting results for each set of experimental data are given in Table~\ref{tab:chsq}, where the details of $\chi^{2} / dof$ for the combined data  (Com. Dat.) are given in the last line. In fact, there are two general parameters for all the models, $\alpha$ and $\kappa$ appearing in $P(s)$, see Eq.~\eqref{eq:ps}. For the Omn\'es model, there is no extra parameter. In the one of HL, there are three more,  $a,\ b,\ c$, as discussed above. Besides, the other BW types, including BW1, BW2, GS and KS models, have another two parameters for the $\rho$ meson, $M_{\rho}$ and $\Gamma_{\rho}$. From the results of Table~\ref{tab:chsq}, one can see that $\chi^{2} / dof$ for Orsay1969 data are all too small due to only a few data points, see Fig.~\ref{fig:gs1}, and the ones for KLOE2005 are much larger than the others owing to the fitted discrepancy around the $\rho-\omega$ mixing region. 
With the summarized results in Table~\ref{tab:chsq} and a systematic analysis of all the fitting results, one can easily find that the results with the GS and KS models are better than the others, which are compatible with each other. 
Indeed, the Omn\'es model has only two free parameters, but it needs the $P$-wave phase shift as input, which is constrained by analyticity and unitarity, as well as crossing symmetry, and depends on the accuracy of the measured $\pi\pi$ $P$-wave phase shift. The HL model is in fact used $N/D$ method for the elastic $P$-wave $\pi\pi$ scattering amplitude with a simple one-pole contribution without considering the detail of  the pole width, which can be matched with the simple BW pole ansatz, see the results later. For the BW1 model, it is only a simple BW pole ansatz from the vector meson dominance, which violates the charge normalization condition as discussed in the last section and is improved by the BW2 model with an energy dependent decay width. Utilized the unitarity condition and considered the detail of the energy dependence of the resonance width for the $\rho$ propagator, this is done in the GS and KS models. Therefore, it is not surprising that the fitting results of the GS and KS models are the best ones.
Thus, our final results are favoured with the one of GS (or KS) model with the fit of the Com. Dat.. In Figs.~\ref{fig:gs1} and \ref{fig:gs2}, we show the results fitted with GS model for each set of experimental data and the Com. Dat., respectively. Note that some data above 1 GeV in the sets of DM1-1978, OLYA1985 and BABAR has been ignored. 

\begin{table}[ht]  
\caption{Results of $\chi^{2} / dof$ for each fit in different sets of experimental data.} 
\label{tab:chsq}
\setlength{\tabcolsep}{4mm}{
 \begin{tabular}{lcccccc} 
  \hline\hline 
Data set   &Omn\'es &HL&BW1&BW2&GS&KS\\
  \hline 
Orsay1969 \cite{Augustin:1969kn}      &0.63&- \footnote{This is due to $dof=0$ with only five data points available.}
&0.02&0.02&0.01&0.01\\
DM1-1978  \cite{Quenzer:1978qt}    &0.74&1.99&0.96&0.92&0.81&0.81\\
OLYA1985 \cite{Barkov:1985ac} &0.57&0.59&0.54&0.54&0.58&0.58\\
CMD1985 \cite{Barkov:1985ac}  &1.72&1.31&1.92&1.78&1.68&1.68\\
CMD2-2002 \cite{Akhmetshin:2001ig}  &1.14&1.32&1.11&1.11&1.14&1.14\\
CMD2-2004\cite{Akhmetshin:2003zn}  &1.14&1.20&1.15&1.14&1.17&1.17\\
CMD2-2006\cite{Akhmetshin:2006wh}  &1.43&12.72&1.7&1.72&1.77&1.79\\
CMD2-2007 \cite{Akhmetshin:2006bx} &2.33&2.71&2.13&2.04&1.86&1.86\\
BaBar2009\cite{Aubert:2009ad}  &1.83&1.06&1.34&1.08&1.05&1.05\\
BESIII2020 \cite{Ablikim:2015orh,Ablikim:2020bah} &1.05&1.10&0.84&0.86&0.95&0.95\\
KLOE2005\cite{Aloisio:2004bu}   &68.42&19.18&21.34&20.52&18.84&18.84\\
KLOE2009 \cite{Ambrosino:2008aa}  &4.74&2.28&6.92&5.3&2.24&2.24\\
KLOE2011 \cite{Ambrosino:2010bv}     &1.09&1.09&1.31&1.15&1.08&1.08\\
KLOE2013 \cite{Babusci:2012rp}      &1.27&1.13&1.53&1.35&1.11&1.11\\
KLOE2018 \cite{Anastasi:2017eio} &1.26&0.77&2.08&1.53&0.77&0.77\\
SND2005 \cite{Achasov:2005rg}  &4.05&3.52&3.54&3.45&3.43&3.43\\
SND2006  \cite{Achasov:2006vp}  &4.04&3.61&3.38&3.36&3.52&3.52\\
SND2020  \cite{Achasov:2020iys} &3.55&3.93&3.39&3.43&3.68&3.68\\
Com. Dat.&11.29&10.20&11.01&10.65&10.19&10.19\\
$\frac{\chi^{2}}{dof}$ (Com. Dat.)
&
$\frac{11337.26}{1006-2}$
&$\frac{10214.46}{1006-5}$
&$\frac{11032.14}{1006-4}$
&$\frac{10671.72}{1006-4}$
&$\frac{10214.59}{1006-4}$
&$\frac{10214.59}{1006-4}$\\
 \hline 
 \end{tabular} }
\end{table}

Furthermore, using the BW types of BW1, BW2, GS and KS models, one can also determine the parameters of $\rho$ meson, $M_{\rho}$ and $\Gamma_{\rho}$, from the fits, see the results of Table~\ref{tab:fitmrho}. Note that in the first column, the ones for the HL model are not obtained directly from the fits. As discussed in Refs.~\cite{Heyn:1980bh,Gardner:1997ie}, once the parameters $a,\ b,\ c$ were determined from the fit, the $M_{\rho}$ and $\Gamma_{\rho}$ could also be determined, since the function $f(s)$, see Eq.~\eqref{eq:fs}, should be matched with the BW form when $s\to M_\rho^2$. And thus, one can have $\mathrm{Re} [f(M_\rho^2)] = 0$ and $\mathrm{Im} [f(M_\rho^2)] = - M_\rho \Gamma_\rho$. Since the small uncertainties are obtained from the fits for the parameters of $a$, $b$, and $c$, the uncertainties for their results of determining $M_{\rho}$ and $\Gamma_{\rho}$ are small too, also for the results of $a_{\mu}^{\mathrm{HVP}, \mathrm{LO}}$ later. For the data of CMD2-2006, because the data is close to the $\pi\pi$ threshold, it is not possible to get the results for $M_{\rho}$ and $\Gamma_{\rho}$ correctly. The results for Orsay1969 are bigger than the others, whereas the ones for KLOE2005 are much smaller. And compared with different models, the results with the BW2 model are bigger than the ones with the BW1 model, and also larger than the numbers obtained with the GS (or KS) model. Finally, we obtain the results of the $\rho$ meson parameters, $M_{\rho}$ and $\Gamma_{\rho}$ 
\footnote{In fact, these results are for the neutral one of $\rho^0$ meson in the $e^+ e^-$ annihilation processes.},
from the fit with GS (or KS) model and the Com. Dat.,
\begin{equation}
M_{\rho} = (774.07\pm0.04) \ \text{MeV}, \quad \Gamma_{\rho} = (149.54\pm0.08) \ \text{MeV},
\end{equation}
which are 1 MeV smaller than the one reported in PDG~\cite{Zyla:2020zbs}, $ (775.26\pm0.23)$ MeV, for the mass, and 2 MeV bigger than the one $ (147.4\pm0.8)$ MeV for the width, and a bit smaller than the other one $M_{\rho} = (775.13\pm0.02)$ MeV obtained in Ref.~\cite{Pich:2001pj}. Our results are also consistent with the one obtained in Ref.~\cite{Davier:2019can}, $M_{\rho} = (774.5\pm0.8)$ MeV. Note that, the small errors for the pole parameters are due to more constraints in the Com. Dat.. One thing should be remarked that, as discussed above and shown in Table~\ref{tab:fitmrho}, the $\rho$ meson parameters, $M_{\rho}$ and $\Gamma_{\rho}$, are model dependent and do not correspond to the physical resonance's mass and width, which should be looked for the pole in the second Riemann sheet. It is complicated to extrapolate the form factor to the second Riemann sheet~\cite{GBarton}, which is out of our concern in the present work and where one can refer to Refs.~\cite{Gonzalez-Solis:2019iod,GomezDumm:2013sib} for more discussions
\footnote{In their model, a pole $(762.0\pm 0.3, \frac{i}{2} (143.0\pm 0.2)\,)$ MeV was found for a mass parameter $M_\rho = (775.2\pm 0.4)$ MeV from the fitting of $\tau$ decay data.} 
and Ref.~\cite{LHCb:2021auc} for the application in the unitarised three-body BW function.

Besides, for the other parameters, we show the results in Table~\ref{tab:par1} in Appendix~\ref{sec:app}, where the details for $\alpha$ and $\kappa$ are given and the results with Omn\'es model are consistent with the ones of Ref.~\cite{Hanhart:2016pcd} within the uncertainties. As done in Ref.~\cite{Hanhart:2016pcd}, one can extract the $\omega\pi\pi$ coupling and the branching ratio of $\omega\to\pi\pi$ from the results of $\kappa$ parameter in Table~\ref{tab:par1}. Thus, following the same way and using the favoured results of GS (or KS) model for the Com. Dat., $\kappa=(1.77\pm0.01) \times 10^{-3}$, we have
\begin{equation}
g_{\omega\pi\pi}=(3.02\pm 0.05) \times 10^{-2}, \quad \text{Br}(\omega\to\pi\pi)=(1.52\pm 0.06) \%,
\end{equation}
which are consistent with the ones obtained in Ref.~\cite{Hanhart:2016pcd} within the uncertainties. Our result for the branching ratio is also in good agreement with the one in PDG~\cite{Zyla:2020zbs}, $\text{Br}(\omega\to\pi\pi)=(1.53^{+0.11}_{-0.13}) \%$.

\begin{table}[ht] 
\caption{Results for the $\rho$ meson parameters $M_{\rho}$ and $\Gamma_{\rho}$ from the fits.}
\label{tab:fitmrho} 
\setlength{\tabcolsep}{2mm}{
 \begin{tabular}{lccccl}  
 \hline\hline  
Data set&HL&BW1&BW2&GS&KS\\
  \hline
Orsay1969&\tabincell{l}{$779.00\pm0.00$\\$158.47\pm0.00$}&\tabincell{l}{$762.10\pm8.99$\\$178.58\pm32.87$}&\tabincell{l}{$793.08\pm13.13$\\$193.37\pm40.43$}&\tabincell{l}{$790.76\pm12.59$\\$190.88\pm38.85$}&\tabincell{l}{$790.72\pm12.59$\\$190.59\pm38.78$}\\
\hline
DM1-1978&\tabincell{l}{$767.53\pm0.00$\\$155.10\pm0.00$}&\tabincell{l}{$757.41\pm3.52$\\$144.63\pm5.11$}&\tabincell{l}{$778.05\pm3.44$\\$153.03\pm6.01$}&\tabincell{l}{$776.83\pm3.49$\\$152.50\pm5.83$}&\tabincell{l}{$776.83\pm3.49$\\$152.50\pm5.83$}\\
\hline
OLYA1985&\tabincell{l}{$775.81\pm5.75$\\$157.54\pm1.69$}&\tabincell{l}{$757.41\pm0.87$\\$143.43\pm1.32$}&\tabincell{l}{$777.54\pm0.77$\\$151.72\pm1.56$}&\tabincell{l}{$775.87\pm0.78$\\$151.71\pm1.53$}&\tabincell{l}{$775.87\pm0.78$\\$151.71\pm1.53$}\\
\hline
CMD1985&\tabincell{l}{$761.57\pm102.54$\\$153.34\pm30.26$}&\tabincell{l}{$756.54\pm2.09$\\$134.54\pm6.51$}&\tabincell{l}{$774.29\pm3.33$\\$140.22\pm7.19$}&\tabincell{l}{$772.47\pm3.15$\\$136.14\pm6.75$}&\tabincell{l}{$772.47\pm3.14$\\$136.12\pm6.73$}\\
\hline
CMD2-2002&\tabincell{l}{$775.38\pm0.00$\\$157.41\pm0.00$}&\tabincell{l}{$758.39\pm0.63$\\$140.96\pm1.22$}&\tabincell{l}{$777.83\pm0.62$\\$148.65\pm1.43$}&\tabincell{l}{$776.40\pm0.62$\\$148.17\pm1.39$}&\tabincell{l}{$776.40\pm0.62$\\$148.17\pm1.39$}\\
\hline
CMD2-2004&\tabincell{l}{$776.02\pm4.92$\\$157.60\pm1.45$}&\tabincell{l}{$758.14\pm0.62$\\$140.78\pm1.22$}&\tabincell{l}{$777.53\pm0.61$\\$148.43\pm1.42$}&\tabincell{l}{$776.07\pm0.62$\\$147.90\pm1.38$}&\tabincell{l}{$776.07\pm0.62$\\$147.90\pm1.38$}\\
\hline
CMD2-2006&$-$&$-$&$-$&$-$&$-$\\
\hline
CMD2-2007&\tabincell{l}{$775.70\pm0.00$\\$157.50\pm0.00$}&\tabincell{l}{$757.82\pm0.44$\\$143.26\pm0.57$}&\tabincell{l}{$777.91\pm0.44$\\$151.46\pm0.68$}&\tabincell{l}{$776.35\pm0.44$\\$151.21\pm0.66$}&\tabincell{l}{$776.35\pm0.44$\\$151.21\pm0.66$}\\
\hline
BaBar2009&\tabincell{l}{$774.53\pm0.00$\\$157.16\pm0.00$}&\tabincell{l}{$755.14\pm0.11$\\$144.60\pm0.24$}&\tabincell{l}{$775.85\pm0.11$\\$152.70\pm0.28$}&\tabincell{l}{$774.58\pm0.11$\\$151.38\pm0.27$}&\tabincell{l}{$774.58\pm0.11$\\$151.38\pm0.27$}\\
\hline
BESIII2020&\tabincell{l}{$775.86\pm0.00$\\$157.55\pm0.00$}&\tabincell{l}{$757.87\pm0.36$\\$142.52\pm1.05$}&\tabincell{l}{$777.73\pm0.40$\\$149.93\pm1.21$}&\tabincell{l}{$776.22\pm0.39$\\$147.87\pm1.17$}&\tabincell{l}{$776.22\pm0.39$\\$147.87\pm1.17$}\\
\hline
KLOE2005&\tabincell{l}{$768.79\pm2.09$\\$155.47\pm0.62$}&\tabincell{l}{$750.52\pm0.13$\\$141.37\pm0.19$}&\tabincell{l}{$770.43\pm0.12$\\$149.63\pm0.22$}&\tabincell{l}{$768.87\pm0.12$\\$149.67\pm0.22$}&\tabincell{l}{$768.87\pm0.12$\\$149.67\pm0.22$}\\
\hline
KLOE2009&\tabincell{l}{$774.86\pm2.06$\\$157.26\pm0.61$}&\tabincell{l}{$756.58\pm0.09$\\$142.07\pm0.13$}&\tabincell{l}{$776.41\pm0.08$\\$150.17\pm0.15$}&\tabincell{l}{$774.93\pm0.08$\\$150.18\pm0.14$}&\tabincell{l}{$774.93\pm0.08$\\$150.18\pm0.14$}\\
\hline
KLOE2011&\tabincell{l}{$775.61\pm0.00$\\$157.48\pm0.00$}&\tabincell{l}{$757.25\pm0.19$\\$142.24\pm0.32$}&\tabincell{l}{$777.11\pm0.16$\\$150.09\pm0.38$}&\tabincell{l}{$775.66\pm0.17$\\$149.29\pm0.37$}&\tabincell{l}{$775.66\pm0.17$\\$149.29\pm0.37$}\\
\hline
KLOE2013&\tabincell{l}{$775.69\pm2.38$\\$157.50\pm0.70$}&\tabincell{l}{$757.30\pm0.28$\\$141.34\pm0.41$}&\tabincell{l}{$776.99\pm0.26$\\$149.40\pm0.49$}&\tabincell{l}{$775.76\pm0.27$\\$149.86\pm0.48$}&\tabincell{l}{$775.76\pm0.27$\\$149.86\pm0.48$}\\
\hline
KLOE2018&\tabincell{l}{$775.15\pm2.67$\\$157.34\pm0.78$}&\tabincell{l}{$756.63\pm0.14$\\$142.29\pm0.20$}&\tabincell{l}{$776.57\pm0.12$\\$150.37\pm0.24$}&\tabincell{l}{$775.22\pm0.13$\\$150.21\pm0.23$}&\tabincell{l}{$775.22\pm0.13$\\$150.21\pm0.23$}\\
\hline
SND2005&\tabincell{l}{$775.27\pm0.00$\\$157.38\pm0.00$}&\tabincell{l}{$756.46\pm0.26$\\$144.34\pm0.49$}&\tabincell{l}{$776.92\pm0.27$\\$152.62\pm0.58$}&\tabincell{l}{$775.33\pm0.27$\\$151.88\pm0.56$}&\tabincell{l}{$775.33\pm0.27$\\$151.88\pm0.56$}\\
\hline
SND2006&\tabincell{l}{$775.54\pm1.92$\\$157.46\pm0.56$}&\tabincell{l}{$756.70\pm0.27$\\$144.27\pm0.49$}&\tabincell{l}{$777.14\pm0.27$\\$152.55\pm0.57$}&\tabincell{l}{$775.60\pm0.27$\\$151.91\pm0.56$}&\tabincell{l}{$775.60\pm0.27$\\$151.91\pm0.56$}\\
\hline
SND2020&\tabincell{l}{$775.59\pm0.00$\\$157.47\pm0.00$}&\tabincell{l}{$757.33\pm0.29$\\$144.33\pm0.81$}&\tabincell{l}{$777.67\pm0.33$\\$152.08\pm0.94$}&\tabincell{l}{$775.92\pm0.32$\\$149.98\pm0.90$}&\tabincell{l}{$775.92\pm0.32$\\$149.98\pm0.90$}\\
\hline
Com. Dat.&\tabincell{l}{$774.00\pm0.79$\\$157.00\pm0.23$}&\tabincell{l}{$755.62\pm0.05$\\$141.74\pm0.07$}&\tabincell{l}{$775.44\pm0.04$\\$149.76\pm0.08$}&\tabincell{l}{$774.07\pm0.04$\\$149.54\pm0.08$}&\tabincell{l}{$774.07\pm0.04$\\$149.54\pm0.08$}\\
\hline
 \end{tabular} }
\end{table}

\begin{figure}[htbp] 
\centering
\includegraphics[scale=1]{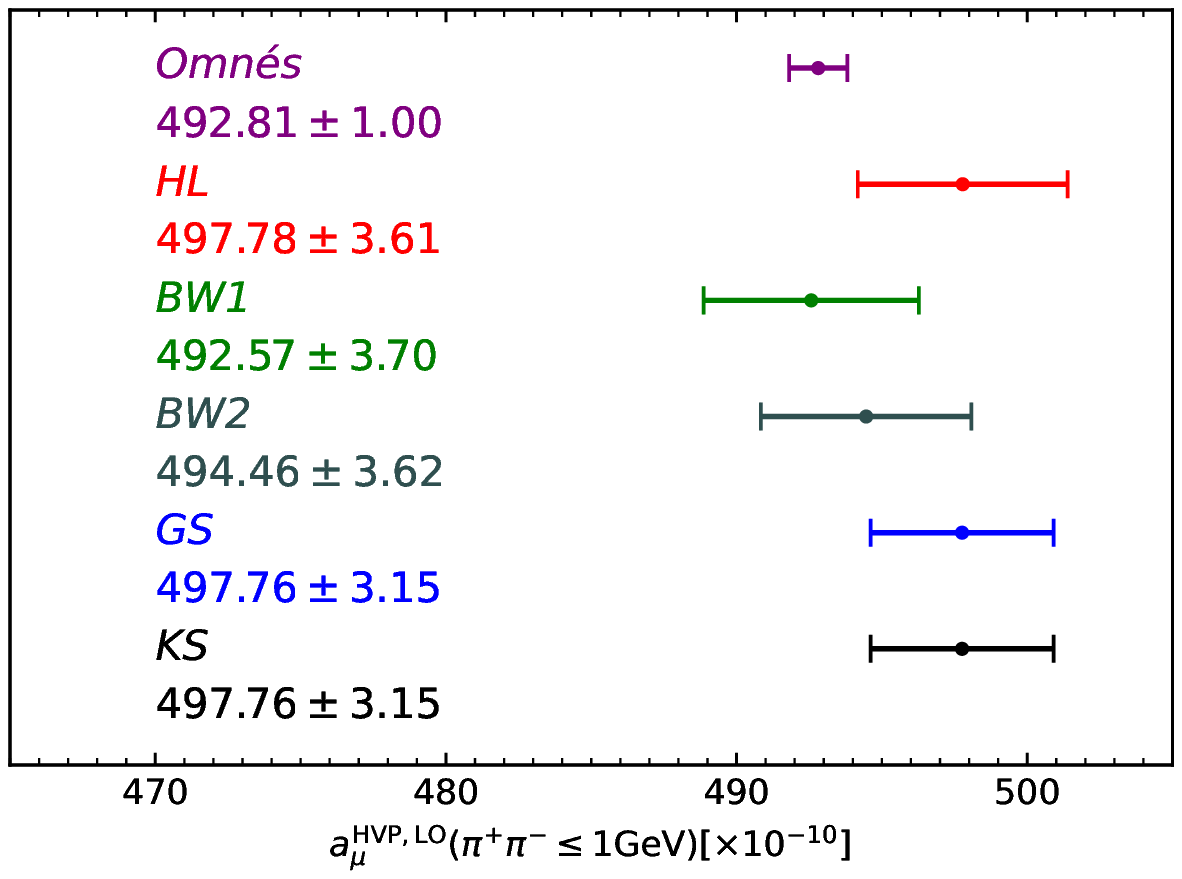}
\caption{Summarized results of $a_{\mu}^{\mathrm{HVP,\; LO}}(\pi^{+} \pi^{-})$ from two-pion contribution up to 1 GeV using the GS model and fitting with the Com. Dat..}
\label{fig:amu}
\end{figure}

With the fitting results obtained above with different models of the PVFF for the data below 1 GeV, we can evaluate the value of $a_{\mu}^{\mathrm{HVP,\; LO}}(\pi^{+} \pi^{-})$ from the two-pion contribution with the dispersion integral defined in Eq.~\eqref{eq:amupi}. Our results are given in Table~\ref{tab:amu}, using different PVFF models for each set of experimental data up to 1 GeV, except for the one of CMD2-2006. We also show the ones with GS model for the Com. Dat. in Fig.~\ref{fig:amu} clearly. As one can find in Table~\ref{tab:amu} that the results for Orsay1969 are bigger than the others, and conversely the ones for CMD1985 are smaller than the others. The uncertainties for the results of Omn\'es model are smaller than the others, whereas the ones with BW1 model are the biggest, see Fig.~\ref{fig:amu}. Indeed, from Fig.~\ref{fig:amu}, compared to the one with GS or KS model, there is a 1\% difference from the smallest one with the BW1 model. At the end, our final results are taken from the one with the GS (or KS) model using the Com. Dat., given as
\begin{equation}
a_{\mu}^{\mathrm{HVP,\; LO}}(\pi^{+} \pi^{-} \leq 1\ \text{GeV}) = (497.76\pm3.15) \times 10^{-10}, 
\end{equation}
which are consistent with the one obtained in Ref.~\cite{Colangelo:2020lcg} very well, $a_{\mu}^{\mathrm{HVP}} (\pi\pi \leq 1\ \text{GeV}) = (497.0 \pm 1.4) \times 10^{-10}$, an updated result of Ref.~\cite{Colangelo:2018mtw} by considering the inelastic effects from the constraints of the Eidelman-{\L}ukaszuk bound. Furthermore, using the framework of resonance chiral theory, two similar values of $a_{\mu}^{\mathrm{HVP}} (\pi\pi \leq 1\ \text{GeV}) = (498.48\pm2.34) \times 10^{-10}$ (Fit I) and $(498.47\pm2.33) \times 10^{-10}$ (Fit II) were obtained in Ref.~\cite{Qin:2020udp}, which are consistent with ours within the uncertainties. One thing should be mentioned that the obtained error is estimated by the average of the reasonable ones in Table~\ref{tab:amu} for different sets of data, since one can find that the errors for different sets of data are mainly contributed by the errors of the pole parameters, see Table~\ref{tab:fitmrho}, and the errors for four sets of data are quite large due to the fewer data points in the $\rho$ region. This is why the error is so small for the one of the Omn\'es model with no pole parameter, see the results of Table~\ref{tab:amu}.

\begin{table}[ht] 
\caption{Results of $a_{\mu}^{\mathrm{HVP,\; LO}}(\pi^{+} \pi^{-})$ ($\times 10^{-10}$) from two-pion contribution up to 1 GeV.} 
\label{tab:amu}
\setlength{\tabcolsep}{1mm}{
 \begin{tabular}{lcccccl} 
  \hline\hline 
Data set&Omn\'es &HL&BW1&BW2&GS&KS\\
  \hline  
Orsay1969&$555.48\pm29.98$&$553.24\pm0.16$&$562.66\pm174.98$&$565.56\pm165.17$&$568.00\pm139.17$&$567.72\pm138.87$\\
DM1-1978&$468.24\pm8.97$&$491.35\pm0.01$&$463.46\pm30.03$&$465.12\pm29.31$&$467.90\pm25.56$&$467.90\pm25.59$\\
OLYA1985&$490.73\pm2.32$&$490.64\pm9.98$&$483.72\pm7.80$&$486.07\pm7.68$&$490.64\pm6.68$&$490.64\pm6.69$\\
CMD1985&$475.48\pm6.12$&$461.50\pm76.32$&$470.32\pm34.68$&$469.71\pm32.50$&$467.19\pm27.65$&$467.20\pm27.64$\\
CMD2-2002&$492.46\pm2.06$&$495.36\pm0.01$&$483.93\pm7.31$&$485.80\pm7.15$&$489.02\pm6.24$&$489.02\pm6.23$\\
CMD2-2004&$503.86\pm2.11$&$501.27\pm9.02$&$496.16\pm7.49$&$498.06\pm7.33$&$501.26\pm6.40$&$501.26\pm6.41$\\
CMD2-2006  &$-$&$-$&$-$&$-$&$-$&$-$\\
CMD2-2007&$501.45\pm0.99$&$501.58\pm0.01$&$494.63\pm3.86$&$496.76\pm3.80$&$500.57\pm3.36$&$500.57\pm3.36$\\
BaBar2009&$501.38\pm0.41$&$503.35\pm0.01$&$500.01\pm1.28$&$501.45\pm1.24$&$503.36\pm1.06$&$503.36\pm1.06$\\
BESIII2020&$498.39\pm1.56$&$500.35\pm0.01$&$493.96\pm5.82$&$495.35\pm5.63$&$496.93\pm4.83$&$496.93\pm4.83$\\
KLOE2005&$473.67\pm0.32$&$496.27\pm7.30$&$490.14\pm1.08$&$492.27\pm1.06$&$496.25\pm0.92$&$496.25\pm0.92$\\
KLOE2009&$495.06\pm0.22$&$497.11\pm7.44$&$491.06\pm0.72$&$493.19\pm0.71$&$497.09\pm0.62$&$497.09\pm0.61$\\
KLOE2011&$494.60\pm0.52$&$493.81\pm0.01$&$489.13\pm1.89$&$490.92\pm1.85$&$493.77\pm1.59$&$493.77\pm1.59$\\
KLOE2013&$495.90\pm0.62$&$495.57\pm6.02$&$490.64\pm2.10$&$492.45\pm2.09$&$495.56\pm1.80$&$495.56\pm1.80$\\
KLOE2018&$495.21\pm0.35$&$495.95\pm7.13$&$490.53\pm1.13$&$492.49\pm1.12$&$495.93\pm0.96$&$495.93\pm0.96$\\
SND2005&$508.68\pm0.77$&$513.28\pm0.01$&$508.34\pm3.19$&$510.25\pm3.12$&$513.28\pm2.73$&$513.28\pm2.75$\\
SND2006&$497.75\pm0.76$&$501.18\pm3.61$&$496.16\pm3.07$&$498.05\pm3.01$&$501.17\pm2.63$&$501.17\pm2.65$\\
SND2020&$500.47\pm0.99$&$499.04\pm0.01$&$498.29\pm5.04$&$499.63\pm4.89$&$501.04\pm4.27$&$501.04\pm4.29$\\
Com. Dat.&$492.81\pm1.00$&$497.78\pm3.61$&$492.57\pm3.70$&$494.46\pm3.62$&$497.76\pm3.15$&$497.76\pm3.15$\\
\hline 
 \end{tabular} }
\end{table}

\section{Conclusion}

In the present work, in order to reduce the uncertainties of the calculation of two-pion contribution to the muon anomalous magnetic moment, we try to get the best fit for the two-pion cross sections with several theoretical models of the pion vector form factor, combined with a polynomial description. Since the polynomial description is valid up to 1 GeV, we only take into account all the experimental data below 1 GeV, which is  below the significant inelastic threshold and contributes almost more than 70\% of the hadronic contribution to the muon anomalous magnetic moment. From our results, we find that the fit with the Gounaris-Sakurai (or K\"uhn-Santamaria) model is the best one. From the best fit to the pion vector form factor, one can also extract the branching ratio of $\omega \to \pi\pi$, given by
\begin{align*}
&\text{Br}(\omega\to\pi\pi)=(1.52\pm 0.06) \%,
\end{align*}
which are compatible with the results reported in Particle Data Group. Based on the best fit to data, we calculate the two-pion contribution to the muon anomalous magnetic moment, obtaining
\begin{equation*}
a_{\mu}^{\mathrm{HVP,\; LO}}(\pi^{+} \pi^{-} \leq 1\ \text{GeV}) = (497.76\pm3.15) \times 10^{-10}, 
\end{equation*}
which is in good agreement with the recent theoretical evaluations~\cite{Colangelo:2020lcg,Qin:2020udp}. Our results for two-pion contribution are helpful to pin down the uncertainties of the calculation for the hadronic vacuum polarization contribution to the muon anomalous magnetic moment.

\section*{Acknowledgments}

We thank Profs. Martin Hoferichter, Irinel Caprini, Hidezumi Terazawa, Peter Athron, Yusi Pan, Wen Yin and Nikolay N. Achasov for valuable comments and useful information, and acknowledge the referee for helpful suggestions.

\appendix

\section{The other parameters}
\label{sec:app}

We give the details of some other parameters in Table~\ref{tab:par1}, where the values of $\alpha$ and $\kappa$ are shown for each fit with different PVFF models, and the ones with Omn\'es model are consistent with the results obtained in Ref.~\cite{Hanhart:2016pcd} within the uncertainties.

\begin{table}[ht] 
\caption{Parameters of $\alpha (GeV^{-2}) $ and $\kappa \times 10^{3}$ for each fit.} 
\label{tab:par1}
\setlength{\tabcolsep}{2mm}{
 \begin{tabular}{lcccccc}  
\hline \hline 
Data set&Omn\'es &HL&BW1&BW2&GS&KS\\
\hline
Orsay1969&\tabincell{l}{$0.15\pm0.05$\\$7.68\pm2.75$}&\tabincell{l}{$0.23\pm0.00$\\$9.26\pm0.00$}&\tabincell{l}{$0.73\pm0.32$\\$18.72\pm8.63$}&\tabincell{l}{$0.69\pm0.29$\\$18.52\pm8.33$}&\tabincell{l}{$0.45\pm0.22$\\$16.79\pm7.21$}&\tabincell{l}{$0.45\pm0.22$\\$16.74\pm7.20$}\\
\hline
DM1-1978&\tabincell{l}{$0.01\pm0.02$\\$2.14\pm0.47$}&\tabincell{l}{$0.12\pm0.00$\\$1.64\pm0.00$}&\tabincell{l}{$0.30\pm0.06$\\$2.61\pm0.61$}&\tabincell{l}{$0.28\pm0.06$\\$2.59\pm0.60$}&\tabincell{l}{$0.12\pm0.05$\\$2.41\pm0.55$}&\tabincell{l}{$0.12\pm0.05$\\$2.41\pm0.55$}\\
\hline
OLYA1985&\tabincell{l}{$0.06\pm0.00$\\$1.68\pm0.17$}&\tabincell{l}{$0.16\pm0.01$\\$1.93\pm0.21$}&\tabincell{l}{$0.34\pm0.02$\\$2.06\pm0.23$}&\tabincell{l}{$0.32\pm0.01$\\$2.05\pm0.23$}&\tabincell{l}{$0.17\pm0.01$\\$1.94\pm0.21$}&\tabincell{l}{$0.17\pm0.01$\\$1.94\pm0.21$}\\
\hline
CMD1985&\tabincell{l}{$0.02\pm0.01$\\$2.03\pm0.26$}&\tabincell{l}{$-0.07\pm0.05$\\$1.52\pm0.30$}&\tabincell{l}{$0.23\pm0.06$\\$2.05\pm0.38$}&\tabincell{l}{$0.20\pm0.06$\\$2.02\pm0.37$}&\tabincell{l}{$0.02\pm0.05$\\$1.79\pm0.33$}&\tabincell{l}{$0.02\pm0.05$\\$1.79\pm0.33$}\\
\hline
CMD2-2002&\tabincell{l}{$0.06\pm0.00$\\$1.48\pm0.13$}&\tabincell{l}{$0.15\pm0.00$\\$1.50\pm0.00$}&\tabincell{l}{$0.32\pm0.01$\\$1.72\pm0.16$}&\tabincell{l}{$0.30\pm0.01$\\$1.72\pm0.16$}&\tabincell{l}{$0.14\pm0.01$\\$1.62\pm0.15$}&\tabincell{l}{$0.14\pm0.01$\\$1.62\pm0.15$}\\
\hline
CMD2-2004&\tabincell{l}{$0.08\pm0.00$\\$1.50\pm0.13$}&\tabincell{l}{$0.16\pm0.01$\\$1.60\pm0.15$}&\tabincell{l}{$0.35\pm0.01$\\$1.71\pm0.16$}&\tabincell{l}{$0.33\pm0.01$\\$1.71\pm0.16$}&\tabincell{l}{$0.17\pm0.01$\\$1.60\pm0.15$}&\tabincell{l}{$0.17\pm0.01$\\$1.60\pm0.15$}\\
\hline
CMD2-2006&$-$&$-$&$-$&$-$&$-$&$-$\\
\hline
CMD2-2007&\tabincell{l}{$0.08\pm0.00$\\$1.54\pm0.05$}&\tabincell{l}{$0.17\pm0.00$\\$1.61\pm0.00$}&\tabincell{l}{$0.37\pm0.01$\\$1.87\pm0.07$}&\tabincell{l}{$0.35\pm0.01$\\$1.86\pm0.07$}&\tabincell{l}{$0.19\pm0.01$\\$1.74\pm0.06$}&\tabincell{l}{$0.19\pm0.01$\\$1.74\pm0.06$}\\
\hline
BaBar2009&\tabincell{l}{$0.08\pm0.00$\\$2.16\pm0.03$}&\tabincell{l}{$0.19\pm0.00$\\$2.25\pm0.00$}&\tabincell{l}{$0.39\pm0.00$\\$2.39\pm0.05$}&\tabincell{l}{$0.37\pm0.00$\\$2.39\pm0.05$}&\tabincell{l}{$0.19\pm0.00$\\$2.26\pm0.04$}&\tabincell{l}{$0.19\pm0.00$\\$2.26\pm0.04$}\\
\hline
BESIII2020&\tabincell{l}{$0.07\pm0.00$\\$1.60\pm0.13$}&\tabincell{l}{$0.16\pm0.00$\\$1.60\pm0.00$}&\tabincell{l}{$0.36\pm0.01$\\$1.98\pm0.18$}&\tabincell{l}{$0.33\pm0.01$\\$1.96\pm0.17$}&\tabincell{l}{$0.16\pm0.01$\\$1.82\pm0.16$}&\tabincell{l}{$0.16\pm0.01$\\$1.82\pm0.16$}\\
\hline
KLOE2005&\tabincell{l}{$0.02\pm0.00$\\$1.78\pm0.03$}&\tabincell{l}{$0.16\pm0.00$\\$1.31\pm0.04$}&\tabincell{l}{$0.35\pm0.00$\\$1.36\pm0.04$}&\tabincell{l}{$0.33\pm0.00$\\$1.37\pm0.04$}&\tabincell{l}{$0.17\pm0.00$\\$1.32\pm0.04$}&\tabincell{l}{$0.17\pm0.00$\\$1.32\pm0.04$}\\
\hline
KLOE2009&\tabincell{l}{$0.07\pm0.00$\\$1.66\pm0.02$}&\tabincell{l}{$0.17\pm0.00$\\$1.75\pm0.03$}&\tabincell{l}{$0.35\pm0.00$\\$1.81\pm0.03$}&\tabincell{l}{$0.33\pm0.00$\\$1.82\pm0.03$}&\tabincell{l}{$0.17\pm0.00$\\$1.75\pm0.03$}&\tabincell{l}{$0.17\pm0.00$\\$1.75\pm0.03$}\\
\hline
KLOE2011&\tabincell{l}{$0.06\pm0.00$\\$1.57\pm0.05$}&\tabincell{l}{$0.15\pm0.00$\\$1.73\pm0.00$}&\tabincell{l}{$0.35\pm0.00$\\$1.82\pm0.06$}&\tabincell{l}{$0.32\pm0.00$\\$1.83\pm0.06$}&\tabincell{l}{$0.16\pm0.00$\\$1.74\pm0.06$}&\tabincell{l}{$0.16\pm0.00$\\$1.74\pm0.06$}\\
\hline
KLOE2013&\tabincell{l}{$0.07\pm0.00$\\$1.39\pm0.10$}&\tabincell{l}{$0.16\pm0.00$\\$1.58\pm0.11$}&\tabincell{l}{$0.34\pm0.00$\\$1.57\pm0.12$}&\tabincell{l}{$0.32\pm0.00$\\$1.59\pm0.12$}&\tabincell{l}{$0.17\pm0.00$\\$1.59\pm0.11$}&\tabincell{l}{$0.17\pm0.00$\\$1.59\pm0.11$}\\
\hline
KLOE2018&\tabincell{l}{$0.07\pm0.00$\\$1.58\pm0.04$}&\tabincell{l}{$0.16\pm0.00$\\$1.69\pm0.05$}&\tabincell{l}{$0.35\pm0.00$\\$1.70\pm0.05$}&\tabincell{l}{$0.33\pm0.00$\\$1.72\pm0.05$}&\tabincell{l}{$0.17\pm0.00$\\$1.69\pm0.05$}&\tabincell{l}{$0.17\pm0.00$\\$1.69\pm0.05$}\\
\hline
SND2005&\tabincell{l}{$0.09\pm0.00$\\$1.71\pm0.04$}&\tabincell{l}{$0.21\pm0.00$\\$1.89\pm0.00$}&\tabincell{l}{$0.41\pm0.01$\\$2.06\pm0.05$}&\tabincell{l}{$0.39\pm0.01$\\$2.05\pm0.05$}&\tabincell{l}{$0.22\pm0.01$\\$1.90\pm0.04$}&\tabincell{l}{$0.22\pm0.01$\\$1.90\pm0.04$}\\
\hline
SND2006&\tabincell{l}{$0.07\pm0.00$\\$1.68\pm0.04$}&\tabincell{l}{$0.19\pm0.00$\\$1.87\pm0.04$}&\tabincell{l}{$0.38\pm0.01$\\$2.02\pm0.05$}&\tabincell{l}{$0.36\pm0.01$\\$2.01\pm0.05$}&\tabincell{l}{$0.19\pm0.00$\\$1.87\pm0.04$}&\tabincell{l}{$0.19\pm0.00$\\$1.87\pm0.04$}\\
\hline
SND2020&\tabincell{l}{$0.08\pm0.00$\\$1.74\pm0.05$}&\tabincell{l}{$0.16\pm0.00$\\$1.86\pm0.00$}&\tabincell{l}{$0.39\pm0.01$\\$2.11\pm0.07$}&\tabincell{l}{$0.36\pm0.01$\\$2.08\pm0.07$}&\tabincell{l}{$0.18\pm0.01$\\$1.90\pm0.06$}&\tabincell{l}{$0.18\pm0.01$\\$1.90\pm0.06$}\\
\hline
Com. Dat.&\tabincell{l}{$0.06\pm0.00$\\$1.76\pm0.01$}&\tabincell{l}{$0.16\pm0.00$\\$1.76\pm0.01$}&\tabincell{l}{$0.35\pm0.00$\\$1.84\pm0.01$}&\tabincell{l}{$0.33\pm0.00$\\$1.85\pm0.01$}&\tabincell{l}{$0.17\pm0.00$\\$1.77\pm0.01$}&\tabincell{l}{$0.17\pm0.00$\\$1.77\pm0.01$}\\
\hline 
 \end{tabular} }
 \end{table}


\end{document}